\documentstyle[11pt]{article}

\newcommand{\Bu}{B${}_{u}$}
\newcommand{\cm}{\mbox{см${}^{-1}$}}
\newcommand{\abio}{\mbox{$\alpha$-Bi${}_{2}$O${}_{3}$}}

\newcommand{\ep}{\epsilon}
\newcommand{\degree}{{}^{o}}
\begin{document}
\normalsize
\thispagestyle{empty}
\vspace{1cm}

\begin{center}

\Large{\bf{Extension of the Kramers-Kronig method for polarized infrared
reflectance spectra from the face of low-symmetry crystals}}

\vspace{1cm}

{\it A.B.Kuz'menko, E.A.Tishchenko, A.S.Krechetov}

\vspace{1cm}

{\normalsize P.L.Kapitza Institute for Physical Problems RAS,

Kosygina str., 2, Moscow, 117334, Russia

\

E-mail: abk@kapitza.ras.ru

}

\vspace{1cm}

{\bf Abstract}

\end{center}

An extension of the Kramers-Kronig method for treatment of 
polarized infrared reflectance spectra from the face of low-symmetry
crystals, where directions of principal dielectric axes depend on
frequency, is proposed. It is shown, how to obtain the frequency
dependencies of the complex reflectivity tensor components, using three
reflectance spectra measured for different directions of linear
polarization of the incident wave, when reflected wave is immediately
sent to the detector. The problem is formulated in a form of the system
of integral equations, and effective numerical technique is found for
solving it. The question of the further recovery of the complex
dielectric tensor on the base of the reflectivity tensor, is discussed.
The case of the monoclinic crystals is considered in details. An
example of the application of the extended Kramers-Kronig method to the
reflectance spectra from the (ac)-face of the single-crystal bismuth
oxide with the monoclinic lattice is given.

\newpage

The Kramers-Kronig method (KK) along with the method of dispersional 
analysis (DA) is the most popular technique for treatment of 
reflectance spectra. As is known, it consists using  the Kramers-Kronig 
integral transformations for recovery of frequency dependence of the 
phase $\theta$ of the complex reflectivity $r=\sqrt{R}\exp{i\theta}$ on 
the base of the real reflectivity $R$ measured in wide enough frequency 
range. Applicability of such a procedure follows from the fact that the 
real part of the analytical function $\ln r=(1/2)\ln R + i\theta$, 
depends only on measured function $R(\omega)$, while its imaginary part 
is unknown function $\theta(\omega)$.  So the function $\theta(\omega)$ 
may be found by integration:

\begin{equation}
\theta(\omega)=-\frac{\omega}{\pi}\int_{0}^{\infty}\frac{\ln{R(\xi)}}
{\xi^{2}-\omega^{2}}d\xi\mbox{\ ,}
\end{equation}

\noindent
where integral with singularity must be treated in the sense of 
principal value \begin{footnote}{Singularity may be eliminated by 
substitution of $\ln{R(\xi)}$ in numerator by 
$\ln{R(\xi)}-\ln{R(\omega)}$.}\end{footnote}.

The next step is calculation of optical characteristics of 
matter. The normal incidence of electromagnetic wave to semi-infinite 
homogenious medium from vacuum is usually discussed. In this case the 
complex dielectric function $\ep$ may be calculated by formula:

\begin{equation}
\label{r2e}
\ep=\left(\frac{1-r}{1+r}\right)^{2}\mbox{\ ,}
\end{equation}

\noindent that followes from the known Fresnel formulas.

An important advantage of the KK method (in comparison with DA) is that it 
allows to obtain the complex dielectric function without model 
calculations, provided that reflectance spectra are known in wide 
enough frequency range. 

However, in the above form, the KK method is valid only when 
dielectric function of crystal and reflectivity coefficient can be 
considered as scalars. It is true when direction of electro-magnetic wave 
polarization coincides with one of the principal axes of the 
dielectric tensor $\hat{\ep}$. In this case eigenvalue of tensor 
$\hat{\ep}$ corresponding to particular axis, plays the role of scalar 
$\ep$. This condition may be satisfied in entire spectral interval only 
for crystals with symmetry no lower than orthorombic. In the case of 
low symmetry crystals (with monoclinic or triclinic syngony) the 
dependence of direction of two or even three axes from frequency takes 
place (so called dispersion of dielectric axes). In monoclinic crystals 
only one axis is fixed in respect to crystalographic system of 
coordinates; two other rotate in the perpendicular plane. In triclinic 
crystals all axes have alternating direction. It doesn't allow  to 
direct polarization of the incident radiation so that the vector of 
electric field and one of the principal axes of the dielectric tensor 
would be parallel. Hence the KK method (and the method of DA also) 
needs to be generalized. 

Generalization of the method DA for phonon spectra in monoclinic 
crystals was first proposed by Belousov and Pavinich 
\cite{Belousov,Belousov2}, who introduced for each IR-active 
phonon an additional angular parameter which characterize direction 
of its dipole momentum and expressed tensor $\hat{\ep}$ through $4n+3$ 
parameters, where $n$ - number of phonon modes. For phonon parameters 
recovery several spectra, obtained for different polarizations of 
incident waves,were used . In \cite{Kuzmenko} we applied this method in 
another form. It was shown that it is sufficient to measure three 
different spectra; every additional spectrum is their linear 
combination. We have measured spectra for three angles of polarization 
equidistant at $45\degree$. Both in \cite{Belousov} and in 
\cite{Kuzmenko} the reflectivity was calculated for the most popular 
situation when only one (the incident) wave is polarized. In this paper 
we offer method for the recovery of the complex reflectivity tensor 
starting from the same three spectra, as in the method DA in form 
\cite{Kuzmenko}. This generalized KK method may be applied for any 
symmetry crystals and may be used in situations, when directions of 
principal axes of the dielectric tensor are not known in advance, as is 
in a case of low symmetry crystals.

\section*{Calculation of the reflectivity tensor}

On assumption of medium linear response, the normal reflection of 
electromagnetic wave from the crystal surface is wholely described by 
the complex reflectivity tensor:

\begin{equation}
\label{tensorr}
 \hat{r}=
\left(
 \begin{array}{cc}
 r_{xx} & r_{xz} \\ 
 r_{zx} & r_{zz}  \\
\end{array} 
\right)\mbox{.}
\end{equation}

\noindent that associates complex amplitudes of incident (i) and 
reflected (r) waves:  ${\bf E}_{r}=\hat{r}{\bf E}_{i}$ (for agreement 
with the notation in \cite{Belousov} and \cite{Kuzmenko}, we now assume 
that the wave vector of the incident wave is directed along the axis 
$y$). The components of the complex amplitude may be phase-shifted
that corresponds to the elliptical polarization of wave. We will 
further assume that the medium is nongyrotropic (it means, for 
example, that no magnetic field is applied) and there is no spatial 
dispersion. In this case reflectivity tensor like dielectric tensor is 
symmetrical:  $r_{xz}=r_{zx}$ \cite{LandauLifshitz,Agranovich}. 

Having set the direction of the incident wave polarization by angle 
$\chi$:

\begin{equation}
{\bf E}_{i} = E_{i}
\left(\begin{array}{c}
\cos\chi \\
\sin\chi        
\end{array}
\right) \mbox{,}
\end{equation}

\noindent we will obtain for the measured reflectivity:

\begin{equation}
\label{refl}
R(\chi)=
\frac{|E_{rx}|^{2}+|E_{rz}|^{2}}{|E_{ix}|^{2}+|E_{iz}|^{2}}=
|r_{xx}\cos{\chi}+r_{xz}\sin{\chi}|^2+
  |r_{xz}\cos{\chi}+r_{zz}\sin{\chi}|^2 \mbox{.}
\end{equation}

The problem is to find spectral dependencies of six functions which 
determine the complex tensor $\hat{r}$ using three reflectance 
spectra $R(0\degree)$, $R(45\degree)$ and $R(90\degree)$. The
structure of the right side of expression (\ref{refl}) doesn't allow to 
separate known and unknown functions by taking of logarithm 
as is possible in isotropic situation. Apparently, solving the problem 
of complex reflectivity tensor recovery by simple integration is 
impossible. Nevertheless, the problem may be formulated in terms of integral 
equations as is shown below.

Presence of two parts in (\ref{refl}) result from the fact that the 
reflected wave is generally elliptically polarized. If another 
polarizer (analyzer) were placed on the way of the wave, one of 
adding would be eliminated. But the scheme with only one polarizer is 
more preferable, because of the following reasons. Firstly, analyzer 
decreases intensity of light that falls onto detector and 
diminishes the signal-to-noise ratio.  Secondly, the scheme discussed 
is the most typical for measurements of IR-spectra and often realized 
in commercial IR-spectrometers. At last, for the proposed variant of 
the method KK and for the method DA on monoclinic crystals 
\cite{Kuzmenko} the same input spectra are used. 

Three equations for unknown functions can be obtained if one 
associate them with the measured reflectance spectra by 
formula (\ref{refl}):

\begin{equation}
\label{R00}
R(0\degree)=|r_{xx}|^2+|r_{xz}|^2 \mbox{,}
\end{equation}
\begin{equation}
\label{R45}
2 R(45\degree)=|r_{xx}+r_{xz}|^2+|r_{xz}+r_{zz}|^2 \mbox{,}
\end{equation}
\begin{equation}
\label{R90}
R(90\degree)=|r_{xz}|^2+|r_{zz}|^2 \mbox{.}
\end{equation}

As was mentioned, additional equations for other polarization 
directions won't add any information, because every new spectrum must 
be a combination of first three ones. This statement follows from type 
of dependence of reflectivity from angle $\chi$ in accordance with 
(\ref{refl}).

Three more equations may be obtained from the KK equations for each 
component of the tensor $\hat{r}$:

\begin{equation}
\label{KKxx}
\mbox{arg }r_{xx}(\omega) = -\frac{2\omega}{\pi}\int_{0}^{\infty}
\frac{\ln{|r_{xx}(\xi)|}}{\xi^{2}-\omega^{2}}d\xi\mbox{,}
\end{equation}
\begin{equation}
\label{KKzz}
\mbox{arg }r_{zz}(\omega) = -\frac{2\omega}{\pi}\int_{0}^{\infty}
\frac{\ln{|r_{zz}(\xi)|}}{\xi^{2}-\omega^{2}}d\xi\mbox{,}
\end{equation}
\begin{equation}
\label{KKxz}
\mbox{arg }[1+r_{xz}(\omega)] = -\frac{2\omega}{\pi}\int_{0}^{\infty}
\frac{\ln{|1+r_{xz}(\xi)|}}{\xi^{2}-\omega^{2}}d\xi\mbox{,}
\end{equation}

\noindent where all integrals are implied in the sense of principal 
value. Shift by one in the last equation is introduced because 
$r_{xz}$, unlike $r_{xx}$ and $r_{zz}$, as function of the complex 
frequency may be equal to zero in the upper semiplane. It disturbs 
analiticity of function $\ln r_{xz}$ and results in inapplicability of 
KK transformations. In contrast, function $\ln(1 + r_{xz})$ is 
analytical in upper semiplane, because $|r_{xz}|<1$.

Equations (\ref{R00})-(\ref{R90}) along with (\ref{KKxx})-(\ref{KKxz}) 
constitute complete system of equation, that may be solved numerically. 

\section*{Calculation of the dielectric tensor}

Having obtained frequency dependencies of all components of reflectance 
tensor, we must determine optical characteristics of crystal, 
particularly, the complex dielectric tensor $\hat{\ep}$. 

Let us make general speculations, valid for all types of crystals, 
without any suggestion about orientation of dielectric axes. At first, 
let's remember that tensor $\hat{\ep}$ in the center of Brillouin zone 
(at ${\bf k}\rightarrow 0$) depends on direction of ${\bf k}$. It is 
governed by fact, that if wave polarization vector ${\bf P}$ has 
longtudinal (relatively to ${\bf k}$) component ${\bf P}_{\parallel}$, 
then a macroscopic electric field appears inside crystal ${\bf 
E}_{\mbox{\small macro}} = -4\pi {\bf P}_{\parallel} = -4\pi {\bf k} 
({\bf k}{\bf P})/|{\bf k}|^{2}$, which depend non-analytically from 
${\bf k}$ at ${\bf k}\rightarrow 0$ \cite{BornHuang}. When the incident 
wave is normal to the surface of crystal the wave vector of induced 
polarization waves (polaritons) inside crystal will also be normal to 
the surface, because in accordance to boundary conditions the 
tangential component of ${\bf k}$ must be constant. That's why all 
obtained values of tensor components will correspond to this direction 
of ${\bf k}$.

The knowledge of the reflectivity tensor $\hat{r}$ with dimensions 
$2\times 2$, for one of the crystal faces is not enough to recover
3-dimensional tensor $\hat{\ep}$. Here we can obtain only partial 
information about $\hat{\ep}$. There is frequently used definition of 
the transverse reflectance tensor $\hat{\ep}_{\perp}$, that correlate 
the vector ${\bf D}$ with the transverse component of vector ${\bf E}$: ${\bf 
D} = \hat{\ep}_{\perp}{\bf E}_{\perp}$. Because vector ${\bf D}$ is 
always perpendicular to the surface, the tensor $\hat{\ep}_{\perp}$ 
is two-dimensional. In accordance to boundary conditions, the vector 
${\bf E}_{\perp}$ must be constant on both sides of surface. Hence, 
it is the tensor $\hat{\ep}_{\perp}$ that define reflectivity tensor 
$\hat{r}$. In \cite{Kuzmenko} were derived formulas that show the 
relation among components of dielectric tensor and that of reflectance 
tensor. In the most compact form this relation may be expressed by 
formula:

\begin{equation} 
\label{epmatr}
\hat{\ep}_{\perp} = \left[(\hat{1} - 
\hat{r})(\hat{1} + \hat{r})^{-1}\right]^2\mbox{\ ,} 
\end{equation} 

\noindent that is similar to (\ref{r2e}) ($\hat{1}$ is the unit 
tensor). Thus, in general case the KK method allows to recover tensor 
$\hat{\ep}_{\perp}$. The further analysis depends on the crystal symmetry and 
particular model, that define features of dielectric function. Relations 
between tensors $\hat{\ep}$ and $\hat{\ep}_{\perp}$ are described in details 
in \cite{Agranovich}.

Let us consider in details crystal with monoclinic lattice. In this 
case the IR-radiation reflectance from the monoclinic (ac)-plane doesn't 
reduce to the scalar case, because directions of two dielectric 
axes in this plane are frequency-dependent. Symmetry allows to extract from 
the tensor $\hat{\ep}$ the two-dimensional minor  $\hat{\ep}_{ac}$, that 
determine dielectric properties in this plane. It coinsides with 
the transverse tensor $\hat{\ep}_{\perp}$ and therefore is defined by formula 
(\ref{epmatr}). Only phonon modes \Bu\ that have dipole momentum in 
(ac)-plane contribute to $\hat{\ep}_{ac}$. In the model of polarized Lorenz 
oscillators this tensor equals:

\begin{equation}
\label{lor}
\hat{\ep}_{ac}(\omega)=\hat{\ep}_{ac}^{\infty} + \sum_{j}
\left(
 \begin{array}{cc}
 \cos^2 \theta_{j} & \cos \theta_{j} \sin \theta_{j} \\
 \cos \theta_{j} \sin \theta_{j} & \sin^2 \theta_{j} \\
\end{array} 
\right)
\frac{S_{j}\omega_{j}^2}{\omega_{j}^2 - \omega^2 - i\gamma_{j}\omega}\mbox{,}
\end{equation}
\noindent where $S_{j}$ - oscillator strength, $\omega_{j}$ - 
transverse frequency, $\gamma_{j}$ - damping coefficient, $\theta_{j}$ 
- angle between direction of dipole momentum and the axis $x$, 
$\hat{\ep}_{ac}^{\infty}$ - high frequency dielectric tensor. 
Summation is performed over all \Bu-modes.

It is convenient to use $\ep_{ac} \equiv \mbox{Sp }\hat{\ep}_{ac} 
= \ep_{xx} + \ep_{zz}$ as uniform scalar characteristics for 
dielectric permeabilty in the (ac)-plane. In the model (\ref{lor}) it 
doesn't depend on directions of oscillator dipole momenta, in 
contrast to $\ep_{xx}$ and $\ep_{zz}$ apart:

\begin{equation}
\label{lor1}
\ep_{ac} = 
\mbox{Sp }\hat{\ep}_{ac}^{\infty} + \sum_{j}
\frac{S_{j}\omega_{j}^2}{\omega_{j}^2 - \omega^2 - i\gamma_{j}\omega}\mbox{.}
\end{equation}

Function (\ref{lor1}) correspond to similar dependency of  
dielectric function from phonon mode characteristics in scalar 
case. Also, similar to a well-known rule for scalar case, transverse 
frequencies of modes \Bu\ may be determined as positions of  
maxima on frequency dependency of $\mbox{Im }\ep_{ac}$. This 
rule, of course, is valid if modes doesn't strongly overlap (interval 
between two neighbour frequencies $\omega_{j}$ greater than their 
widths). Under this assumption the polarization angles $\theta_{j}$ (defined 
with ambiguity to $\pi$) may be approximately obtained from equation:

\begin{equation}
\cos 2\theta_{j} \approx \frac 
{\mbox{Im }\ep_{xx}(\omega_{j})-\mbox{Im }\ep_{zz}(\omega_{j})} 
{\mbox{Im }\ep_{xx}(\omega_{j})+\mbox{Im }\ep_{zz}(\omega_{j})} 
\mbox{\ ,} 
\end{equation} \noindent which 
might be drawn out if from imaginary part of sum  (\ref{lor}) 
the contribution of only $j$-th mode (dominating at 
$\omega=\omega_{j}$) is considered. The choice of right root 
is determined by sign of off-diagonal element of $\mbox{Im 
}\ep_{xz}(\omega_{j})$.

\section*{Features of numerical procedure}

Having substituted:
\begin{equation}
a e^{i\alpha} = r_{xx}\mbox{,\ \ }
b e^{i\beta} = r_{zz}\mbox{,\ \ }
c e^{i\gamma} = 1 + r_{xz}\mbox{,\ \ }
\end{equation}
\begin{equation}
A = R(0\degree)\mbox{,\ \ }
B = R(90\degree)\mbox{,\ \ }
C = R(45\degree) - (A+B)/2\mbox{,}
\end{equation}
\noindent we'll find that system of equations  (\ref{R00})-(\ref{KKxz}) 
changes to:

\begin{equation}
\label{A}
A = a^2 + c^2 - 2 с\cos \gamma + 1\mbox{,}
\end{equation}
\begin{equation}
\label{B}
B = b^2 + c^2 - 2 с\cos \gamma + 1\mbox{,}
\end{equation}
\begin{equation}
\label{C}
C = ac\cos(\alpha-\gamma) + bc\cos(\beta-\gamma)  
- a\cos \alpha - b\cos \beta\mbox{,}
\end{equation}
\begin{equation}
\label{alpha}
\alpha(\omega)=-\frac{2\omega}{\pi}\int_{0}^{\infty}\frac{\ln{a(\xi)}}
{\xi^{2}-\omega^{2}}d\xi\mbox{,}
\end{equation}
\begin{equation}
\label{beta}
\beta(\omega)=-\frac{2\omega}{\pi}\int_{0}^{\infty}\frac{\ln{b(\xi)}}
{\xi^{2}-\omega^{2}}d\xi\mbox{,}
\end{equation}
\begin{equation}
\label{gamma}
\gamma(\omega)=-\frac{2\omega}{\pi}\int_{0}^{\infty}\frac{\ln{c(\xi)}}
{\xi^{2}-\omega^{2}}d\xi\mbox{}
\end{equation}

\noindent with six unknown functions $a$, $b$, $c$,
$\alpha$, $\beta$ и $\gamma$ and three known functions $A$, $B$ и $C$.

Let's vary unknown functions so that they satisfy system of 
equations. Inside measured frequency interval we choose quite dense 
grid of base points $\omega_{1} < \omega_{2} < \ldots < 
\omega_{n}$. We approximately specify functions $a$, $b$ и $c$ by they 
values $a_{i}=a(\omega_{i})$, $b_{i}=b(\omega_{i})$ and 
$c_{i}=c(\omega_{i})$, which assume as unknown variables. Then we calculate 
values of functions $\alpha$, $\beta$ и $\gamma$ in base points,  
approximately substituting in formulas  (\ref{alpha})-(\ref{gamma}) 
integration by summation:

\begin{equation} \label{sum} \alpha_{i} = 
\sum_{j=1}^{n}\sigma_{ij}\ln a_{j}\mbox{, \ \ } \beta_{i} = 
\sum_{j=1}^{n}\sigma_{ij}\ln b_{j}\mbox{, \ \ } \gamma_{i} = 
\sum_{j=1}^{n}\sigma_{ij}\ln c_{j}\mbox{, \ \ } i=1,\ldots,n\mbox{\ ,} 
\end{equation} 

\noindent where coefficients $\sigma_{ij}$ are expressed from  
$\omega_{1}, \ldots, \omega_{n}$ according to method of 
approximate integration (type of interpolation function, method of 
extrapolation outside interval). For example, if one substitute a 
partly-constant function instead of real function $f(\omega)$:

\begin{equation}
f^{*}(\omega) = 
\left\{
\begin{array}{cl}
f_{1} & 
\mbox{,\ } \omega < (\omega_{1}+\omega_{2})/2  \\
f_{j} & 
\mbox{,\ } (\omega_{j-1}+\omega_{j})/2 < \omega 
< (\omega_{j}+\omega_{j+1})/2 \mbox{, \ \ } j=2, \ldots, n-1 \\
f_{n} &
\mbox{,\ } \omega > (\omega_{n-1}+\omega_{n})/2  \mbox{\ \ ,}
\end{array} \right. 
\end{equation} 

\noindent the following values of coefficients  $\sigma_{ij}$ will 
be obtained:

\begin{equation}
\sigma_{ij} = -\frac{1}{\pi}\cdot
\left\{
\begin{array}{ll}
\ln 
\frac{\textstyle \left| 2\omega_{i}-\omega_{j}-\omega_{j+1}\right|}
     {\textstyle 2\omega_{i}+\omega_{j}+\omega_{j+1}}
& 
\mbox{,\ } j=1 \\
\ln 
\frac{\textstyle \left|2\omega_{i}-\omega_{j}-\omega_{j+1}\right|}
     {\textstyle 2\omega_{i}+\omega_{j}+\omega_{j+1}}
-
\ln 
\frac{\textstyle \left|2\omega_{i}-\omega_{j-1}-\omega_{j}\right|}
     {\textstyle 2\omega_{i}+\omega_{j-1}+\omega_{j}}
& 
\mbox{,\ } j=2, \ldots, n-1 \\
-
\ln
\frac{\textstyle \left|2\omega_{i}-\omega_{j-1}-\omega_{j}\right|}
     {\textstyle 2\omega_{i}+\omega_{j-1}+\omega_{j}}
& 
\mbox{,\ } j=n\mbox{\ \ .}
\end{array}
\right.
\end{equation} 

It corresponds to extrapolation of low- and high-frequency spectra by 
constant function, that is acceptable for dielectrics. This method 
was used to treating model and real spectra presented in this paper.

In each point there are three measured quantities 
$A_{i}=A(\omega_{i})$, $B_{i}=B(\omega_{i})$ and $C_{i}=C(\omega_{i})$. 
After definition of set of values $a_{i}$, $b_{i}$ and $c_{i}$ and 
calculating with help of (\ref{sum}) $\alpha_{i}$, $\beta_{i}$ and 
$\gamma_{i}$, the following functional can be composed:

\begin{equation}
\label{chisq}
F = \frac{1}{3n}\sum_{i=1}^{n}\left[
\left(\frac{A_{i}-A^{\prime}_{i}}{\delta A_{i}}\right)^2 + 
\left(\frac{B_{i}-B^{\prime}_{i}}{\delta B_{i}}\right)^2 + 
\left(\frac{C_{i}-C^{\prime}_{i}}{\delta C_{i}}\right)^2
\right] \mbox{,}
\end{equation}

\noindent 
where trial quantities $A^{\prime}_{i}$, $B^{\prime}_{i}$ and
$C^{\prime}_{i}$ are expressed through $a_{i}$, $b_{i}$,
$c_{i}$, $\alpha_{i}$, $\beta_{i}$, $\gamma_{i}$ in accordance with
(\ref{A})-(\ref{C}), and  $\delta A_{i}$, $\delta B_{i}$, 
$\delta C_{i}$ are meansquare noise deviations of $A_{i}$, 
$B_{i}$, $C_{i}$.

Let's vary  $3n$ quantities $a_{i}$, $b_{i}$ and $c_{i}$ to 
find a minimum of functional (\ref{chisq}) as function of these 
quantities. As initial approximation it is reasonable to take  
$a_{i} =\sqrt{A_{i}}$, $b_{i} = \sqrt{B_{i}}$, $c_{i} = 1$, what is 
exact decision in a case when the reflectivity tensor (\ref{tensorr}) is 
diagonal. Set of values that makes functional approximateli equal one 
may be considered as approximate decision.

For functional minimization we have used the Marquardt-Levenberg 
nonlinear optimization method described in \cite{NumericalRecipes}.
Partial derivatives of functional by variables  $a_{i}$, $b_{i}$ and 
$c_{i}$ were calculated in analytical form. It is advisable to vary 
the density of base points according to density of spectrum "features"
for diminishing $n$. In our calculations maximum number of base 
points $n$ was 450. Despite great number of variables, minimization 
process converged quickly enough (5-7 iterations).

Certainly, it is not so obvious, why must minimization process 
always lead to true result, because functional F in  
multi-dimensional space may have local minima that do not correspond to 
required solution. Besides, the noise in input spectra may 
cause instability of minimization process. However our experience shows 
that minimum which corresponds to right decision, is usually so deep 
that minimization procedure converges just to it. Experimental 
noise (in reasonable limits) doesn't affect the final result very much.  
Nevertheless, the convergence of method and its stability to 
experimental noises require special attention.

\section*{ Application examples }

Before application of the discussed method KK to real experimental 
data, we tested it on a set of model spectra obtained with a help of 
the dispersion formulas for monoclinic lattices \cite{Kuzmenko}, 
taking different sets of parameters of \Bu\ phonon modes 
(frequencies, intensities, widths and polarization directions). To examine 
stability of the KK method to noise, additional noise with amplitude 5\% 
was applied to initial spectra. Calculated by the KK method spectra for 
all components of the reflectivity tensor and the dielectric tensor were in 
a good agreement with that obtained directly by DA formulas. On the 
figure 1 results of one such test are showed.

Application of the discussed method to real reflectance 
spectra of single crystal \abio, that has monoclinic lattice is shown as an 
illustration. On the figure 2(a) three reflectance spectra from the face (ac) 
are presented. They were used as input data for the generalized KK 
method. Details of experiment and results of data treatment by the DA method  
are described in \cite{Kuzmenko}. 

Calculated by the KK method imaginary part of dielectric function  
$\ep_{ac}$ of bismuth oxide is presented on the figure 2(b) along 
with its frequency dependence calculated by the DA method. In general, a good 
match between these data may be noted. Several deviations, however, 
demonstrate that common description of dielectric function by Lorentz 
oscillators, isn't quite adeqiate. Inasmuch as discussion of optic 
properties of bismuth oxide isn't topic of this paper, we restrict ourselves 
by this notice.

\ 
 
Authors are thankful to E.L.Kosarev, A.I.Kleev and E.R.Podolyak for 
discussion of problems concerning solving of system of integral equations.

\newpage

{\bf Figure captures}
\ 

\

{\bf  Fig.1.} Results of the generalized KK method application to the model 
problem about reflection of IR-radiation from the (ac)-face of 
monoclinic crystal with two polarized in the (ac)-plane Lorentz oscillators, 
with characteristics: $\omega_{1}=200\mbox{\ } \cm$, $S_{1}=0.56$, 
$\gamma_{1}=10\mbox{\ }\cm$, $\theta_{1}=30\degree$; $\omega_{2}=450\mbox{\ } 
\cm$, $S_{2}=0.79$, $\gamma_{2}=20\mbox{\ }\cm$, $\theta_{2}=60\degree$.  
Components $\hat{\ep}_{ac}^{\infty}$:  
$\ep_{xx}=\ep_{zz}=5.0$, 
$\ep_{xz}=0.0$. 

(a) Reflectance spectra at different polarization angles $\chi$, obtained  
according to formulas of dispersional analysis with added artificial noise 
5\%. 1 - $0\degree$ (shifted at 1), 2 - $45\degree$ (shifted at 0.5), 3 
- $90\degree$.

(b) Components of reflectance tensor $\hat{r}$, obtained  as a results of 
treatment spectra shown on figure 1a with help of KK method (dots) and 
directly with help of the DA formulas DA without additional noise (solid 
line).  1 - $\mbox{Im }r_{xx}$, 2 - $\mbox{Re }r_{xx}$, 3 - $\mbox{Im 
}r_{zz}$ (shifted at -1), 4 - $\mbox{Re }r_{zz}$ (shifted at -1), 5 - 
$\mbox{Im }r_{xz}$ (shifted at -2), 6 - $\mbox{Re }r_{xz}$ (shifted at  
-2.5).

(c) The imaginary part of the dielectric function $\ep_{ac}$, calculated
on the base of reflectivity tensor (figure 2a). Dots - reflectivity  
tensor was recovered using the KK method. Solid line - "true" reflectivity 
tensor calculated directly by the DA formulas. 

\

{\bf Fig.2.} 

(a) IR reflectance spectra at $T$=300 K from the (ac) face of 
monoclinic crystal \abio\ at different polarization angles $\chi$.  1 - 
$0\degree$ (shifted at 2), 2 - $45\degree$ (shifted at  1), 3 - 
$90\degree$.

(b) Frequency dependence $\mbox{Im }\ep_{ac}$, recovered by two methods 
from initial spectra drawn at figure 2a. Dots - function values obtained 
with a help of the generalized KK method. Solid line - results obtained 
with a help of the DA method \cite{Kuzmenko}.


\newpage

\begin{thebibliography}{100}

\bibitem{Jahoda}{{\it Jahoda F.C.}, Phys.Rev., 1957, v.107, n.5, p. 
1261-1265.}

\bibitem{Spitzer}{{\it Spitzer W.G., Kleinmann D.A.}, Phys.Rev., 1961, 
v.121, n.5, p.1324-1335.}

\bibitem{Belousov}{{\it Belousov M.V., Pavinich V.}, Optics and 
Spectroscopy (Russian), 1978, v.45, n.5, p.920-926.}

\bibitem{Belousov2}{{\it Belousov M.V., Pavinich V.}, Optics and 
Spectroscopy (Russian), 1978, v.45. n.6, p.1114-1118.}

\bibitem{Kuzmenko}{{\it Kuz'menko A.B., Tishchenko E.A., Orlov V.G.}, 
J. Phys.: Condens. Matter, 1996, v.8, p.6199-6212.}

\bibitem{LandauLifshitz}{{\it Landau L.D., Lifshitz E.M.},
Electrodynamics of Continous Media, 1975, Oxford, Pergamon}

\bibitem{Agranovich}{{\it Agranovich V.M., Ginzburg V.L.},
Crystal optics with spatial disperdion and excitons, 1984, v.42, 
Berlin, Springer}

\bibitem{BornHuang}{{\it Born B. and Huang K.}, Dynamical 
theory of crystal lattices. Oxford: Clarendon Press, 1954}

\bibitem{NumericalRecipes}{{\it Press W.H., Teulkolsky S.A., Vetterling 
W.T., Flannery B.P.}, Numerical Recipes in FORTRAN. Cambridge:  
Cambridge University Press, 1992}

\end{thebibliography}
\end{document}